\documentclass[journal,10pt,onecolumn,]{IEEEtran}
\usepackage{amsmath,amsfonts,amssymb,amsthm,epsfig}
\usepackage{graphicx}
\usepackage{caption}
\usepackage{subcaption}
\usepackage{enumerate}
\usepackage{hyperref}
\usepackage{url}
\usepackage{mathtools, cuted}
\usepackage{mathrsfs}
\usepackage{amsbsy}

\usepackage{bbold}
\newtheorem{theorem}{Theorem}

\newtheorem{lemma}{\textit{Lemma}}

\def\x{{\mathbf x}}
\def\X{{\mathbf X}}

\def\Y{{\mathbf Y}}
\def\A{{\mathbf A}}
\def\B{{\mathbf B}}
\def\D{{\mathbf D}}
\def\U{{\mathbf U}}

\def\O{{\Omega}}

\begin{document}

\title{Tensor Matched Kronecker-Structured Subspace Detection for Missing Information}

\author{\IEEEauthorblockN{Ishan Jindal, \emph{Student Member, IEEE} and 
Matthew Nokleby, \emph{Member, IEEE}}\\
\IEEEauthorblockA{Department of Electrical and Computer Engineering,
Wayne State University,
Detroit, MI, 48202,  USA\\ Email: \{ishan.jindal, matthew.nokleby\}@wayne.edu}
% \thanks{This paper in part was presented at the 2017 IEEE Symposium on Information Theory \cite{jindal2017performance} and at the 2017 Asilomar Conference on Signals, Systems, and Computers \cite{jindal2017fast}}
}

% make the title area
\maketitle

\begin{abstract}
We consider the problem of detecting whether a tensor signal having many missing entities lies within a given low dimensional Kronecker-Structured (KS) subspace. This is a matched subspace detection problem. Tensor matched subspace detection problem is more challenging because of the intertwined signal dimensions. We solve this problem by projecting the signal onto the Kronecker structured subspace, which is a Kronecker product of different subspaces corresponding to each signal dimension. Under this framework, we define the KS subspaces and the orthogonal projection of signal onto the KS subspace. We prove that the reliable detection is possible as long as the cardinality of missing signal is greater than the dimensions of the KS subspace by bounding the residual energy of the sampling signal with high probability.
\end{abstract}

% no keywords

% For peer review papers, you can put extra information on the cover
% page as needed:
% \ifCLASSOPTIONpeerreview
% \begin{center} \bfseries EDICS Category: 3-BBND \end{center}
% \fi
%
% For peerreview papers, this IEEEtran command inserts a page break and
% creates the second title. It will be ignored for other modes.
%\IEEEpeerreviewmaketitle

%!Tex root = Tensor_journal.tex
\section{Introduction}
\label{sec:intro}

The matched subspace detection problem arises in different scientific areas. This includes medical imaging such as in image representation \cite{li2014transformation}, in shape detection \cite{wang2017shape}, communication MIMO network systems \cite{sarieddeen2016efficient,mccloud2000interference}, matrix completion \cite{candes2009exact,krishnamurthy2013low} etc., where we need to detect whether a given signal lies within a subspace. The matched subspace detection is well studied problem \cite{scharf1994matched,azizyan2012subspace}. However, this problem become more challenging when the signal is not completely observable that is only a small subset of the signal entries are known and based on this observation we want to test whether a signal belongs to a given subspace.  

To deal with subspace detection with missing information, \cite{balzano2010high} and \cite{wu2014subspace} provides a way to deal with missing information in the signal. These available methods make an implicit assumption that the signal is present in vectorized form and convert a multi-dimensional signal into a single dimension before testing. Many real-world signals such as dynamic scene video \cite{gangopadhyay2016dynamic} or tomographic images are inherently multi dimensional, which capture the spatial and temporal correlations within the data. However, by vectorizing the signal we lose the multi-dimensional structure of the data, which could be used to enhance the performance of the detector. Previous works, for example \cite{elad2006image}, consider general subspace structure, whereas our work applies a subspace model to tensors by supposing that each mode of the tensor lies approximately along a subspace similar to \cite{jindal2017performance}. Equivalently, we preserve this multi-dimensional structure of the signal by projecting the signal onto Kronecker-structured subspace, which is a Kronecker product of a number of subspaces corresponding to the dimensionality of the signal, while observing only a small subset of the elements of the signal; hence the K-S subspace model is a special case of general subspace models. Authors in \cite{bahri2018robust} and \cite{jindal2018classification} show how the multi-dimensional structure in data can be well exploited for better classification and representation performance. 
 % Robust Kronecker Component Analysis \cite{bahri2018robust}.

% In this paper, we preserve this multi-dimensional structure of the signal by projecting the signal onto Kronecker-structured subspace, which is a Kronecker product of a number of subspaces corresponding to the dimensionality of the signal, while observing only a small subset of the elements of the signal. 
In this work, we present all our analysis for 2-D signals $\Y \in \mathbf{R}^{m_1\times m_2}$. We study two different ways through which information can be missing, First, when the entire row and/or column of the 2-D signal is missing, represented by $\Y_\O$. For instance, while capturing the EEG signals using number of electrodes placed over the scalp, one of the electrodes is broken or we miss to capture the signal for a time window. Second, a more general case, when only discrete entries are missing, that is, the missing entries need not to be in the form of entire rows and/or columns, represented by $\Y_{\widehat{\O}}$. We formulate a binary hypothesis test for the more general case of missing information, where given a Kronecker-structured subspace $\D \in ~\mathbb{R}^{m_1m_2\times n_1n_2}$ we need to find whether $\Y \in \D$ by just observing the samples of $\Y$ with or without noise. 

We formulate the hypothesis test as $\mathcal{H}_0: \Y \in \D$ and $\mathcal{H}_1: \Y \notin \D$. This test follows immediately by computing the residual energy, that is, if $\Y \in \D$, then the residual energy $\left|\left|\Y - \U^A\Y\U^B\right|\right|_F^2 = 0$ and $\left|\left|\Y_\O - \U_\O^A\Y_\O\U_\O^B\right|\right|_F^2 = 0$, where $\U^A$ and $\U^B$ are the orthogonal projection operator onto row and column subspace, respectively. We show that the residual energies of the signal are bounded with high probability. The main result of this work answers, given the row and column subspaces with dimensions $n_1 \ll m_1$ and $n_2 \ll m_2$, respectively, how many rows and columns of the 2-D signal must be observed in order to reliably detect whether the signal belongs to the given subspace.

The rest of the paper is organized as follows: In Section \ref{sec:MD}, we define the Kronecker structured subspace model. We present the final results in Section \ref{sec:mainRes} with the discussion and experiments  in Section \ref{sec:experiment}. The hypothesis testing for matched subspace detection is provided in Section \ref{sec:msd} and finally, we conclude this work with future directions in Section \ref{sec:conclusion}. 
% For all the proofs, in this paper, we refer the reader to \cite{arxiv_version}.

\textit{\textbf{Mathematical Notation:}} We use lowercase bold letters to represent vectors, such as $\x$ and use uppercase bold to represent matrices, such as $\Y$. We use $\otimes$ to represent the Kronecker product between two matrices. 
%!Tex root = ./Tensor_journal.tex

\section{Model Definition}
\label{sec:MD}

A two dimensional signal of interest $\Y \in \, \mathbb{R}^{m_1 \times m_2}$ is represented by a matrix $\A \in \mathbb{R}^{m_1 \times n_1}$, which describes the subspace on which the columns of $\Y$ approximately lie with $n_1 \leq m_1$, and by $\B \in \, \mathbb{R}^{m_2 \times n_2}$, which describes the subspace on which the rows of $\Y$ approximately lie with $n_2\leq m_2$, that is $\Y = \A\X\B^T$ where $\X \in \mathbb{R}^{n_1 \times n_2}$ represents the coefficient matrix. We can also express $\Y$ in vectorized form
\begin{equation}
\mathbf{y} = (\A \otimes \B)\x,
\end{equation}
 where $\mathbf{y} = \text{vec}(\Y)$, $\mathbf{x} = \text{vec}(\X)$ and $\otimes$ denotes the Kronecker product.
% For a tensor classification model, $\Y$ is the set of observation and . $\A $, $\B \, $, , $\D = \A \otimes \B \, \in \, \mathbb{R}^{m_1m_2 \times n_1n_2 }$. Where $\Y = \A\X\B^T$, $n_1 \leq m_1$, $n_2\leq m_2$ and $\A, \B$ are the row and column subspace where the signal $\Y$ lies approximately.
Similar to \cite{balzano2010high}, we define the \emph{coherence} of Kronecker-structured subspace ~${\D = \A~\otimes \B ~\in ~\mathbb{R}^{m_1m_2\times n_1n_2}}$ as 
\begin{align}
\mu(\D) := \frac{m_1m_2}{n_1n_2} \underset{j}{\text{max}} \left|\left|\U^{\D} e^{AB}_j\right|\right|^2_2,
\label{eq:coherence}
\end{align}
where $\U^D = \D\left(\D^T\D\right)^{-1}\D^T$ is the orthogonal projection operator onto $\D$, $e_j^{AB}$ represents the standard basis vector and $||\cdot||_2$ represents the euclidean norm of a vector. The coherence provides the amount of information we can expect from each sample to provide. From \cite{candes2009exact}, the coherence can take values between ~$1\leq ~\mu(\D) \leq \frac{m_1m_2}{n_1n_2}$~. For further analysis, we also need to know whether the signal has energy outside the $n_1$ and $n_2$ dimensional column and row subspaces, respectively. Therefore, we define the coherence of column and row subspaces as
% require this amount of information we can expect for each of the subspace to provide as the coherence of the subspace $\A \otimes \B$.
\begin{align}
\mu(\A) &:= \frac{m_1}{n_1} \underset{j}{\text{max}} \left|\left|\U^A e^A_j\right|\right|^2_2,\\ \mu(\B) &:= \frac{m_2}{n_2} \underset{j}{\text{max}} \left|\left|\U^B e^B_j\right|\right|^2_2.
\end{align}
Where, $\U^A = \A\left(\A^T\A\right)^{-1}\A^T$ and $\U^B = \B\left(\B^T\B\right)^{-1}\B^T$ are the projection operators onto $\A$ and $\B$ subspace, respectively and can take values ~$1\leq ~\mu(\A) \leq \frac{m_1}{n_1}$~ and ~$1\leq ~\mu(\B) \leq \frac{m_2}{n_2}$~. 
To establish the relationship between the coherence of Kronecker-structured subspace with the coherence of individual subspaces we need the Lemma \ref{lemma:coherence}.
\begin{lemma}
The coherence of the Kronecker-structured subspace is equal to the product of coherence of individual subspaces, that is 
\begin{equation}
\mu(\A \otimes \B) = \mu(\A) \mu(\B).
\label{eq:coherenceKS}
\end{equation} 
\label{lemma:coherence}
\end{lemma}
\begin{proof}
in Appendix \ref{Appendix:coherence}
\end{proof}

$\mu(\A)$ achieves minimum values when all the vectors whose all the entries have magnitude $\frac{1}{\sqrt{n_1}}$ forms $\A$ and if $\A$ contains a standard basis element then $\mu(\A)$ achieves the maximum value $\frac{m_1}{n_1}$. Similar analysis holds for $\mu(\B)$ and $\mu(\D)$. For a 2-D signal $\Y$, we let $\mu(\Y)$ defines the coherence of the subspace spanned by the signal $\Y$ and we define the $l_{\infty}$ norm as $||\Y||_{\infty} = \underset{{i,j}}{\max } \,\,\text{abs}(\Y(i,j))$. From \cite{candes2009exact}, we write $\mu(\Y) = m_1m_2 \frac{|| \Y||_{\infty}^2}{|\\Y||_F^2}$.
% \cite{candes2009exact} states the definition of coherence of subspace $\U$ of dimension r in $\mathbb{R}^n$ with orthogonal projector $\P_U = \U(\U^T\U)^{-1}\U^T$ onto $\U$ with the standard basis $e_i$ is defined as:
% $$\mu(\U) := \frac{n}{r} \underset{j}{max} ||\P_U e_j||^2_2.$$ Here, $1 \leq \mu(U) \leq \frac{n}{r}$. $\mu(U) = 1$ when all the vectors whose all the entries have magnitude $\frac{1}{\sqrt{n}}$ forms $U$. and if $U$ contains a standard basis element then the maximum value of $\mu(U) = n/r$.

\subsection{Missing Signals}
\begin{figure}[!ht]
\vspace{-0.1in}
\centering
	% \begin{subfigure}{0.3\columnwidth}
	% \centering
	% 	\includegraphics[width = \textwidth]{./images/signalY.png}
	% 	\caption{$\Y$}
	% 	\label{fig::signal}
	% \end{subfigure}
	\begin{subfigure}{0.45\columnwidth}
	\centering
		\includegraphics[width = 0.9\textwidth]{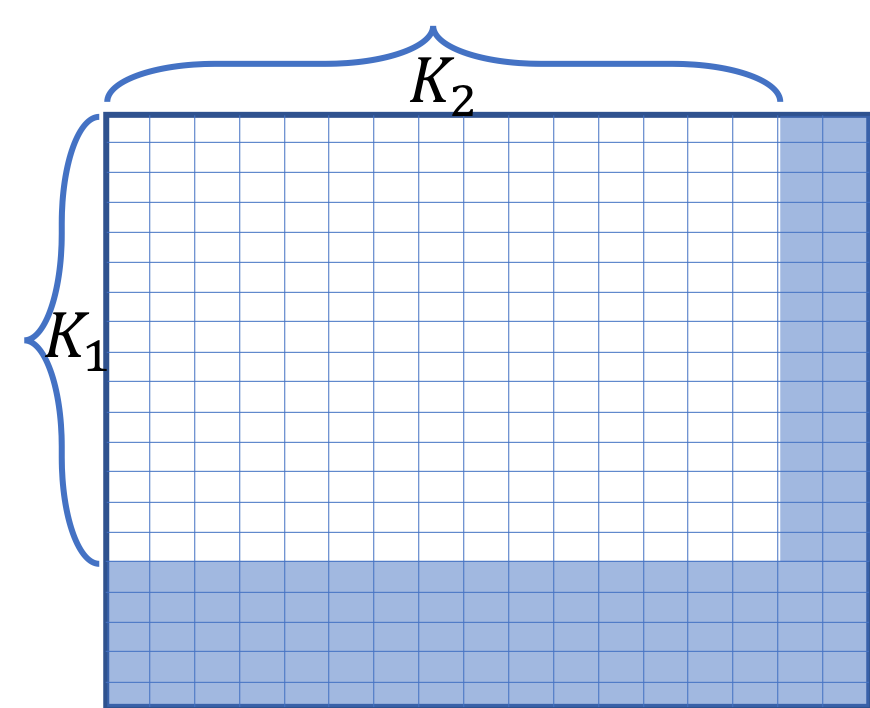}
		\caption{$\Y_{\Omega_{k_1k_2}}$}
		\label{fig::Inter}
	\end{subfigure}
	\begin{subfigure}{0.45\columnwidth}
	\centering
		\includegraphics[width = 0.9\textwidth]{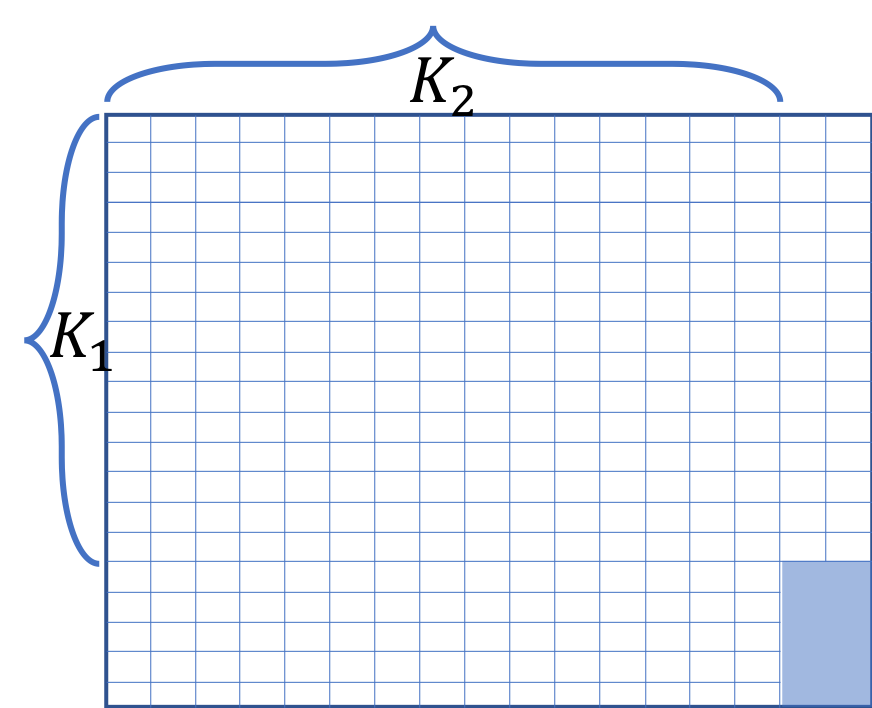}
		\caption{$\Y_{\widehat{\O}(k_1,k_2)}$}
		\label{fig::Union}
	\end{subfigure}
	\caption{(a) Intersection and (b) union of rows and columns.}
	\label{fig:UnionInter}
\end{figure}

For tensor signals, we can expect signal entries to be missing along one or many dimensions, for instance, in Fig. \ref{fig::Inter} entire rows and columns of the 2-D signal $\Y$ are missing. We represent the signal with this type of missing information as $\Y_\O$, where $\Omega_{k_1k_2}$ represents the index of non-zero rows $(k_1)$ and non-zeros columns $(k_2)$. Now onwards, we use shorthand $\Omega$ in replace of $\Omega_{k_1k_2}$ throughout the paper. Thus, $\Y_\O$ is a signal of dimension $k_1 \times k_2$. The energy of the signal $\Y$ in the subspace $\A\otimes \B$ is $\left|\left|\U^A\Y(\U^B)^T\right|\right|_F^2$, where $\U^A$ and $\U^B$ are the column and row projection operator onto $\A$ and $\B$ subspaces, respectively and $||\cdot||_F$ represents the Frobenius norm of the matrix. 

Now, we define the column and row subspace for missing signal as $\A_\O \in  \mathbf{R}^{k_1\times n_1}$ and $\B_\O \in  \mathbf{R}^{k_2\times n_2}$. Here, $k_1$ and $k_2$ are the columns and rows of $\A$ and $\B$, respectively indexed by the set $\Omega_{k_1k_2}$, arranged in lexigraphic order. Since we only observe the signal $\Y$ for the set of rows and columns indexed by $\Omega_{k_1k_2}$, we estimate the missing signal energy $\Y_\O$ in $\D$ as how well the missing signal is best represented by the subspace $\D_\O = \A_\O\otimes \B_\O$ with the projection operator $\U_\O^D = \D_\O\left(\D_\O^T\D_\O\right)^{-1}\D_\O^T$. Therefore, if the row and columns of signal $\Y$ lies in row and column subspace then $\left|\left|\Y - \U^A\Y\U^B\right|\right|_F^2 = 0$ and hence $$\left|\left|\Y_\O - \U_\O^A\Y_\O\U_\O^B\right|\right|_F^2 = \left|\left|\Y_\O - \widehat{\Y}_\O\right|\right|_F^2 =  0,$$ where $\U_\O^A = \A_\O\left(\A_\O^T\A_\O\right)^{-1}\A_\O^T$ and $\U_\O^B = \B_\O\left(\B_\O^T\B_\O\right)^{-1}\B_\O^T$. 

However, it is not always true that the entire row or/and entire column of the signal is missing. It may be the case that, while collecting the sensor output, some sensors are dead for a period of time and then wake up again. Therefore, we extend our analysis to a more general case of missing information, that is when any particular entry in the signal is missing as shown in Fig. 1(b) where only a fraction of entries are missing. We represent the signal with missing discrete entries as $\Y_{\widehat{\O}(k_1,k_2)}$, where $\widehat{\Omega}(k_1,k_2) = \{(i, j) \forall (i,j) | \Y_{\widehat{\O}(i,j)} \neq 0\}$ represents the location of all the non-zero entries.
% , that is, $(k_1, k_2) = \{(i, j) \forall (i,j) \Big| \Y_{\widehat{\O}(i,j)} \neq 0\}$.

When the entire rows and/or columns of the signal are missing, we represent the remaining signal as the \emph{intersection} of remaining rows and columns. Whereas, when discrete entries are missing, we represent the remaining signal as the signal minus the intersection of rows and/or columns that contains missing entries, we call it as \emph{union} of dimensions. For further clarification, in Fig. \ref{fig:UnionInter} we plot a 2-D signal $\Y \in \mathbf{R}^{20 \times 17}$ with $m_1 = 20$ and $m_2 = 17$. Let 5 rows and 2 columns of the signal $\Y$ is missing, we represent the remaining signal as the intersection of  $k_1 = m_1 - 5 = 15$ and $ k_2 = m_2-2 = 15$ in Fig. \ref{fig::Inter}. Similarly, for the missing discrete entries we count the number of rows and columns to which the missing entries belongs and subtract the count from corresponding signal dimensions to obtain $k_1$ and $k_2$. Finally subtract the intersection of the $k_1$ and $k_2$ from the signal to represent the remaining signal as shown in Fig. \ref{fig::Union}.

% Here we deal with a special case when either the entire row, or column or both are missing from the 2-D signal $Y$ and represent it as $Y_{\Omega}$, where $\Omega(i,j)$ represent the index of rows $(i)$ and columns $(j)$ which are non zero. Therefore for the missing entry signal we have the projection operators as $U^A_{\Omega} = A_{\Omega}(A_{\Omega}^TA_{\Omega})^{-1}A_{\Omega}^T$ and $U^B_{\Omega} = B_{\Omega}(B_{\Omega}^TB_{\Omega})^{-1}B_{\Omega}^T$ and the reconstructed signal can be written as:$$ \hat{Y}_{\Omega} = U^A_{\Omega}Y_{\Omega}U^B_{\Omega}.$$

%!Tex root = Tensor_journal.tex

\section{Main Results}
\label{sec:mainRes}

\begin{theorem}
Let the entire rows or/and columns of the signal be missing, $\delta > 0$, $k_1 \geq \frac{8}{3}n_1\mu(\A)\log\left(\frac{2n_1}{\delta}\right) $ and $k_2\geq \frac{8}{3}n_2\mu(\B)\log\left(\frac{2n_2}{\delta}\right)$, than with probability at least $1-8\delta$,
\begin{multline}
\Bigg((1-\alpha)\left(\frac{k_1k_2}{m_1m_2}\right) - \frac{(\beta+1)^2}{2(1-\gamma_2)(1-\gamma_1)} \Big( \frac{n_1}{m_1} \mu(\A) +  \frac{n_2}{m_2} \mu(\B)\Big) \Bigg) ||\Y- \U^A\Y\U^B||^2_F \leq ||\Y_{\Omega} - \U^A_{\Omega}\Y_{\Omega}\U^B_{\Omega} ||_F^2 \\ \leq (1+\alpha)\left(\frac{k_1k_2}{m_1m_2}\right)\left|\left|\Y- \U^A\Y\U^B\right|\right|_F^2,
\end{multline}
\label{th:intersection}
where $\alpha = \sqrt{\frac{2\mu(\Y)^2}{k_1k_2}\log(\frac{1}{\delta})}$, $\beta = \sqrt{\frac{4\mu(\Y)\log(\frac{1}{\delta})}{\left(\frac{m_2}{n_2}\frac{1}{\mu(\B)}+\frac{m_1}{n_2}\frac{1}{\mu(\A)}\right)} }$, $\gamma_1 = \sqrt{\frac{8n_1\mu(\A)}{3k_1}\log\left(\frac{2n_1}{\delta}\right)}$ and $\gamma_2 = \sqrt{\frac{8n_2\mu(\A)}{3k_2}\log\left(\frac{2n_2}{\delta}\right)}$.
\end{theorem}

\begin{proof}
To prove this theorem, we first solve for the residual energy as follows
\begin{align}
||\Y_{\Omega} - \U^A_{\Omega}\Y_{\Omega}\U^B_{\Omega} ||_F^2  &=  \text{tr}\left(\left(\Y_{\Omega} -  \U^A_{\Omega}\Y_{\Omega}\U^B_{\Omega}\right)^T\left(\Y_{\Omega} -  \U^A_{\Omega}\Y_{\Omega}\U^B_{\Omega}\right)\right)\nonumber \\
&= \left|\left|\Y_{\Omega} \right|\right|_F^2 - \text{tr}\left(\Y_{\Omega}^T\U^A_{\Omega}\Y_{\Omega}\U^B_{\Omega}\right) \nonumber \\
& = \left|\left|\Y_{\Omega} \right|\right|_F^2 -  \text{tr}\left(\Y_{\Omega}^T \A_{\Omega}\left(\A_{\Omega}^T\A_{\Omega}\right)^{-1}\A_{\Omega}^T \Y_{\Omega} \B_{\Omega}\left(\B_{\Omega}^T\B_{\Omega}\right)^{-1}\B_{\Omega}^T\right).
\label{eq::residual}
\end{align}
Now to solve the second part in \eqref{eq::residual}, define $\mathbf{T}_{\Omega}^T\mathbf{T}_{\Omega} = \left(\A_{\Omega}^T\A_{\Omega}\right)^{-1} $ and $\mathbf{V}_{\Omega}\mathbf{V}_{\Omega}^T = \left(\B_{\Omega}^T\B_{\Omega}\right)^{-1}$. Using the rotation property of trace we write
\begin{equation}
\text{tr}\left(\Y_{\Omega}^T\A_{\Omega}\mathbf{T}_{\Omega}^T\mathbf{T}_{\Omega}\A_{\Omega}^T \Y_{\Omega} \B_{\Omega} \mathbf{V}_{\Omega}\mathbf{V}_{\Omega}^T\B_{\Omega}^T\right) \\= \left|\left|\mathbf{T}_{\Omega} \A_{\Omega}^T \Y_{\Omega} \B_{\Omega} \mathbf{V}_{\Omega}^T\right|\right|_F^2.
\end{equation}
% the II term in \eqref{eq::residual} as
% \begin{align*}
% \left|\left|\mathbf{T}_{\Omega} \A_{\Omega}^T \Y_{\Omega} \B_{\Omega} \mathbf{V}_{\Omega}^T\right|\right|_F^2 &= \text{tr}\left(\mathbf{V}_{\Omega}^T\B_{\Omega}^T\Y_{\Omega}^T\A_{\Omega}\mathbf{T}_{\Omega}^T\mathbf{T}_{\Omega}\A_{\Omega}^T \Y_{\Omega} \B_{\Omega} \mathbf{V}_{\Omega}\right) \\
% & = \text{tr}\left(\Y_{\Omega}^T\A_{\Omega}\mathbf{T}_{\Omega}^T\mathbf{T}_{\Omega}\A_{\Omega}^T \Y_{\Omega} \B_{\Omega} \mathbf{V}_{\Omega}\mathbf{V}_{\Omega}^T\B_{\Omega}^T\right)
% \end{align*}
Substituting $\mathbf{T}_{\Omega}^T\mathbf{T}_{\Omega}$ and $\mathbf{V}_{\Omega}\mathbf{V}_{\Omega}^T$ we obtain \eqref{eq::residual} as:
\begin{align}
\left|\left|\Y_{\Omega} - \widehat{\Y}_{\Omega} \right|\right|_F^2 &= \left|\left|\Y_{\Omega} \right|\right|_F^2 - \left|\left|\mathbf{T}_{\Omega} \A_{\Omega}^T \Y_{\Omega} \B_{\Omega} \mathbf{V}_{\Omega}^T\right|\right|_F^2\nonumber \\
&\leq \left|\left|\Y_{\Omega} \right|\right|_F^2 - \left|\left|\mathbf{T}_{\Omega}\right|\right|_F^2 \cdot \left|\left|\A_{\Omega}^T \Y_{\Omega} \B_{\Omega}\right|\right|_F^2 \cdot \left|\left|\mathbf{V}_{\Omega}^T\right|\right|_F^2.
% \nonumber \\
% & = \left|\left|\Y_{\Omega} \right|\right|_F^2 - \left|\left|\left(\A_{\Omega}^T\A_{\Omega}\right)^{-1}\right|\right|_F \cdot \left|\left|\A_{\Omega}^T \Y_{\Omega} \B_{\Omega}\right|\right|_F^2 \cdot \left|\left|\left(\B_{\Omega}^T\B_{\Omega}\right)^{-1}\right|\right|_F
\end{align}
The final expression for residual energy is
\begin{equation}
\left|\left|\Y_{\Omega} - \widehat{\Y}_{\Omega} \right|\right|_F^2 \leq \left|\left|\Y_{\Omega}\right|\right|_F^2 -\Bigg( \left|\left|\left(\A_{\Omega}^T\A_{\Omega}\right)^{-1}\right|\right|_F^2 \\ \cdot \left|\left|\A_{\Omega}^T \Y_{\Omega}\B_{\Omega}\right|\right|\cdot \left|\left|\left(\B_{\Omega}^T\B_{\Omega}\right)^{-1}\right|\right|_F^2\Bigg).
\label{eq:final_res}
\end{equation}
We bound the each term in \eqref{eq:final_res} with high probability using the following Lemmas.
 % I: $||Y_{\Omega} ||_F^2$, II: $||\left(\A_{\Omega}^TA_{\Omega}\right)^{-1}||_F$, III: $||A_{\Omega}^T Y_{\Omega} B_{\Omega}||_F^2$ and IV: $||\left(\B_{\Omega}^TB_{\Omega}\right)^{-1}||_F$.
\end{proof}
% Now we need to find the lower bound and upper bound on the projection residual as follows:
% $$ \text{L.B} (H_1) \leq ||Y_{\Omega} - \hat{Y}_{\Omega} ||_F^2 \leq \text{U.B} (H_0).$$
% After applying some maths we obtain:

\begin{lemma}
Using the similar notations as Theorem \ref{th:intersection} we bound the missing signal energy as
\begin{equation}
(1-\alpha)\left(\frac{k_1k_2}{m_1m_2}\right)||\Y||_F^2 \leq ||\Y_{\Omega}||_F^2 \\ \leq (1+\alpha)\left(\frac{k_1k_2}{m_1m_2}\right)||\Y||_F^2.
\label{eq::KSEnBound_inter}
\end{equation}
\label{lm::KSEnBound_inter}
with probability at least $1-2\delta$.
\end{lemma}
\begin{lemma}
With high probability $1-2\delta$ and using same notation in Theorem \ref{th:intersection}
\begin{equation}
||\A_{\Omega}^T \Y_{\Omega} \B_{\Omega}||_F^2 \leq (\beta+1)^2 \\ \frac{1}{2}\frac{k_1k_2}{m_1m_2} \left( \frac{n_1}{m_1} \mu(\A) +  \frac{n_2}{m_2} \mu(\B)\right) ||\Y||^2_F.
\label{eq::CoeffBound_inter}
\end{equation}
\label{lm::CoeffBound_inter}
\end{lemma}
\begin{lemma}
With high probability $1-\delta$ and using same notation in Theorem \ref{th:intersection}, we bound the energy of column subspaces as 
\begin{equation}
 \left|\left|\left(\sum_{i\in \Omega}\A_{\Omega_{i,:}} \A_{\Omega_{i,:}}^T\right)^{-1}\right|\right|_F \leq \frac{m_1}{(1-\gamma_1)k_1}.
 \label{eq::AsubBound_inter}
 \end{equation}
  Similarly, we bound the energy of row subspace as
\begin{equation}
 \left|\left|\left(\sum_{i\in \Omega}\B_{\Omega_{:,i}} \B_{\Omega_{:,i}}^T\right)^{-1}\right|\right|_F \leq \frac{m_2}{(1-\gamma_2)k_2}.
 \label{eq::BsubBound_inter}
 \end{equation}
 \label{lm::ABsubBound_inter}
\end{lemma}
Combining the Lemmas \ref{lm::CoeffBound_inter} and \ref{lm::ABsubBound_inter} with the energy of missing signal in Lemma \ref{lm::KSEnBound_inter} and using the union bound we obtain the final expression in Theorem \ref{th:intersection}.

We further extend this analysis to a more general scenario when discrete data entries are missing, that is for example, rather than missing an entire row and/or column, only some of the entries from that row and/or column are missing as shown in Fig. \ref{fig::Union}.
\begin{theorem}
Let the discrete entries from the signal be missing, $\delta > 0$, $k_1 \geq \frac{8}{3}n_1\mu(\A)\log\left(\frac{2n_1}{\delta}\right) $ and $k_2\geq \frac{8}{3}n_2\mu(\B)\log\left(\frac{2n_2}{\delta}\right)$, than with probability at least $1-8\delta$,
\begin{multline}
\frac{k_1m_2+k_2m_1-k_1k_2}{m_1m_2}\Bigg((1-\alpha) - \frac{m_1m_2(\beta+1)^2}{2k_1k_2(1-\gamma_1)(1-\gamma_2)} \left( \frac{n_1}{m_1} \mu(\A)+  \frac{n_2}{m_2} \mu(\B)\right) \Bigg) \left|\left|\Y- \U^A\Y\U^B\right|\right|^2_F \\ \leq \left|\left|\Y_{\Omega} - \U^A_{\Omega}\Y_{\Omega}\U^B_{\Omega} \right|\right|_F^2  \leq (1+\alpha)\frac{k_1m_2+k_2m_1-k_1k_2}{m_1m_2}||\Y||_F^2,
\end{multline}
\label{th:union}
where $\alpha = \sqrt{\frac{2\mu(\Y)^2k_1k_2}{(k_1m_2+k_2m_1-k_1k_2)^2}\log(\frac{1}{\delta})}$, $\beta = \sqrt{\frac{4\mu(\Y)\log(\frac{1}{\delta})}{\left(\frac{m_1}{k_1}+\frac{m_2}{k_2}-1\right)\left(\frac{m_2}{n_2}\frac{1}{\mu(\B)}+\frac{m_1}{n_2}\frac{1}{\mu(\A)}\right)} }$, $\gamma_1 = \sqrt{\frac{8n_1\mu(\A)}{3k_1}\log\left(\frac{2n_1}{\delta}\right)}$ and $\gamma_2 = \sqrt{\frac{8n_2\mu(\A)}{3k_2}\log\left(\frac{2n_2}{\delta}\right)}$.
\end{theorem}

% \begin{figure}[]
% \centering
% \includegraphics[width = 0.5\textwidth]{./images/Set_inter_Union.png}
% \caption{Intersection of dimensions in terms of Union of dimensions}
% \label{fig::inter_union}
% \end{figure}

\begin{lemma}
Using the similar notations as Theorem \ref{th:union} we bound the missing signal energy as
\begin{equation}
(1-\alpha)\frac{k_1m_2+k_2m_1-k_1k_2}{m_1m_2}\left|\left|\Y\right|\right|_F^2 \leq \left|\left|\Y_{\Omega}\right|\right|_F^2 \\ \leq (1+\alpha)\frac{k_1m_2+k_2m_1-k_1k_2}{m_1m_2}\left|\left|\Y\right|\right|_F^2,
\label{eq::KSEnBound_union}
\end{equation}
\label{lm::KSEnBound_union}
with probability at least $1-2\delta$.
\end{lemma}

\begin{lemma}
With high probability $1-2\delta$ and using same notation in Theorem \ref{th:union}
\begin{equation}
\left|\left|\A_{\Omega}^T \Y_{\Omega} \B_{\Omega}\right|\right|_F^2 \leq (\beta+1)^2\frac{k_1m_2+k_2m_1-k_1k_2}{2m_1m_2} \cdot\\ \left( \frac{n_1}{m_1} \mu(\A)+  \frac{n_2}{m_2} \mu(\B)\right) ||\Y||^2_F.
\label{eq::CoeffBound_union}
\end{equation}
\label{lm::CoeffBound_union}
\end{lemma}
Now we put together Lemmas \ref{lm::KSEnBound_union}, \ref{lm::CoeffBound_union} and \ref{lm::ABsubBound_inter} in \eqref{eq:final_res} and using the union bound to obtain the final Theorem \ref{th:union}.

Theorem \ref{th:intersection} always provides the tighter bound than the Theorem \ref{th:union}. For example, When only one entry from the signal is missing then from the pre-multiplier of upper bound in both the theorems we say that $k_1k_2 < k_1m_2+k_2m_1-k_1k_2$, because in any case $m_1>k_1$ and $m_2>k_2$. Furthermore, both the theorems can be very easily extended to tensors with more than 2 dimensions. For the tensors with more than 2 dimensions, the Kronecker subspace is the Kronecker product of more than two subspaces and rest of the analysis follows immediately.
% \begin{figure}[h!]
% \centering
% 	\begin{subfigure}[t]{0.45\columnwidth}
% 	\centering
% 		\includegraphics[width = 0.5\textwidth]{./images/Union15_15_1.png}
% 		\caption{Show Union of $K_1= 15$ and $K_2 = 15$}
% 		% \label{fig::KSUnion}
% 	\end{subfigure}
% 	\begin{subfigure}[t]{0.45\columnwidth}
% 	\centering
% 		\includegraphics[width = 0.5\textwidth]{./images/Inter15_15.png}
% 		\caption{Show Inter of $K_1= 15$ and $K_2 = 15$}
% 		% \label{fig::KSInter}
% 	\end{subfigure}
% \end{figure}
% \textbf{Shaded region indicates missing pixels}
% \begin{figure}[h!]
% \centering
% 	\begin{subfigure}[t]{0.45\columnwidth}
% 	\centering
% 		\includegraphics[width = 0.5\textwidth]{./images/Union1_1.png}
% 		\caption{Show Union of $K_1= 1$ and $K_2 = 1$}
% 		% \label{fig::KSUnion}
% 	\end{subfigure}
% 	\begin{subfigure}[t]{0.45\columnwidth}
% 	\centering
% 		\includegraphics[width = 0.5\textwidth]{./images/Inter1_1.png}
% 		\caption{Show Inter of $K_1= 1$ and $K_2 = 1$}
% 		% \label{fig::KSInter}
% 	\end{subfigure}
% \end{figure}

% This means that the energy of signal with missing entries is same in both the case when we keep the union or intersection of dimensions.

%!Tex root = Tensor_journal.tex

\section{Analysis and Experiments}
\label{sec:experiment}
\begin{figure}[!h]
\centering
	\begin{subfigure}{0.48\columnwidth}
	\centering
	\includegraphics[width=\columnwidth]{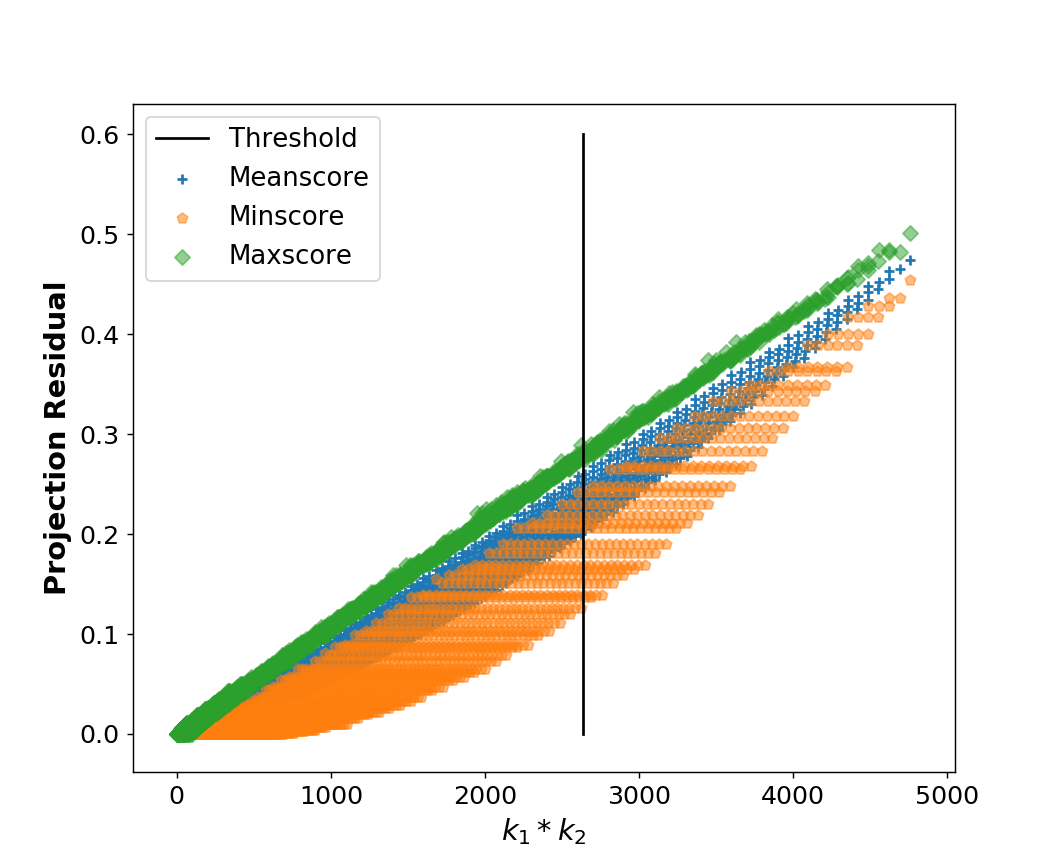}
	\caption{$\Y \in (\A \otimes \B)^{\bot}$}
	\label{fig::ProjectionAB}
	\end{subfigure}
	~
	\begin{subfigure}{0.48\columnwidth}
	\centering
	\includegraphics[width=\columnwidth]{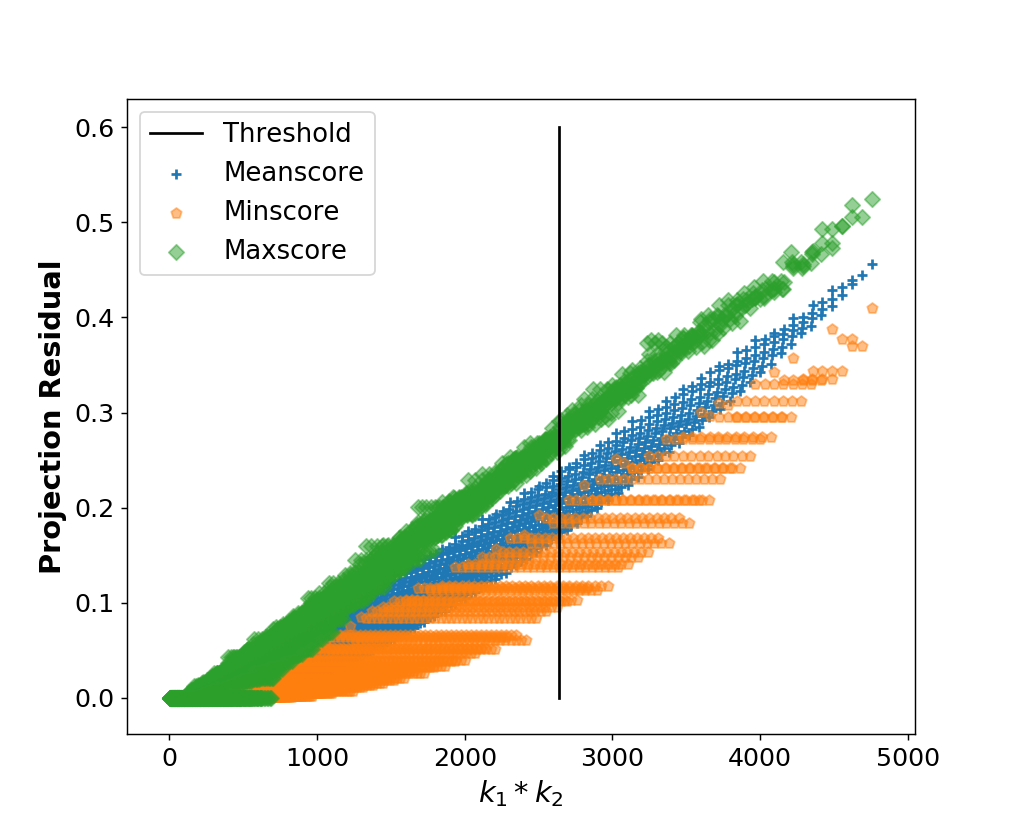}
	\caption{$\Y \in (\A^{\bot} \otimes \B)$}
	\label{fig::projectionA}
	\end{subfigure}

	\begin{subfigure}{0.48\columnwidth}
	\centering
	\includegraphics[width=\columnwidth]{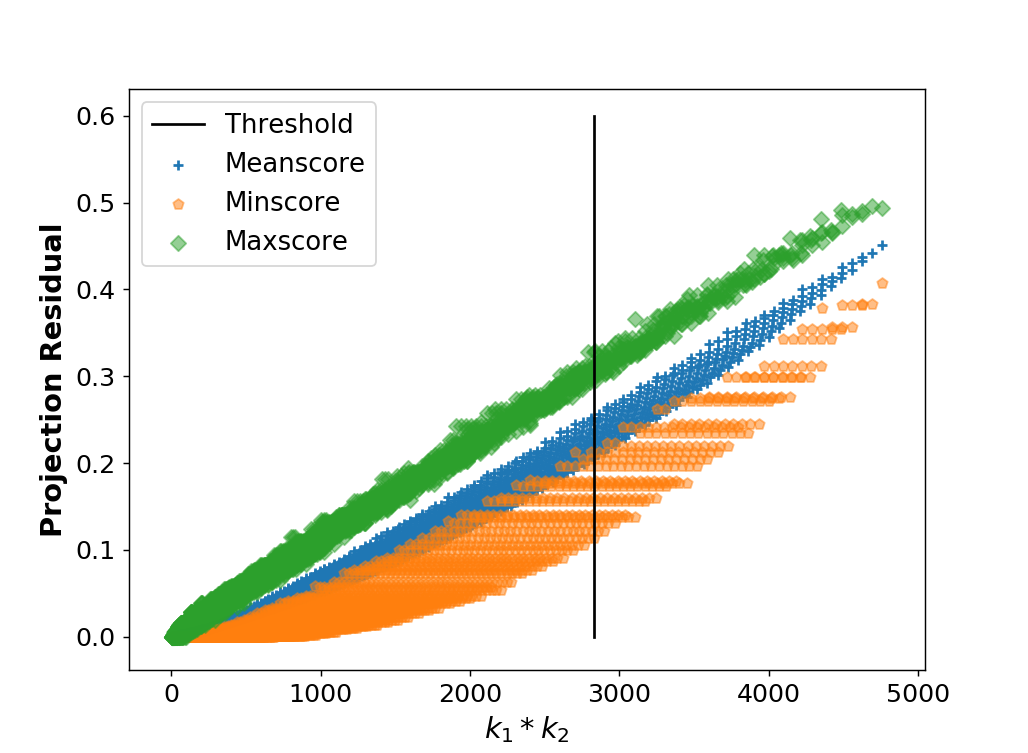}
	\caption{$\Y \in (\A \otimes \B^{\bot})$}
	\label{fig::projectionB}
	\end{subfigure}
	~
	\begin{subfigure}{0.48\columnwidth}
	\centering
	\includegraphics[width=\columnwidth]{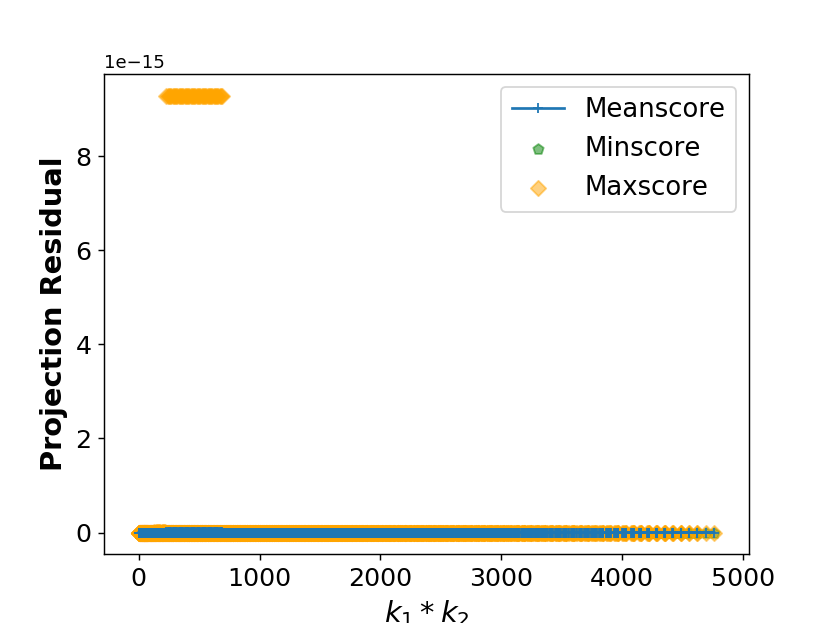}
	\caption{$\Y \in (\A \otimes \B)$}
	\label{fig::ptojectionNO}
	\end{subfigure}
\caption{The projection residual $\left|\left|\Y_{\widehat{\O}} - \U^A_{\widehat{\O}}\Y_{\widehat{\O}}\U^B_{\widehat{\O}} \right|\right|_F^2$ averaged over 1000 simulations for fixed row $\A \in \mathbb{R}^{k_1\times 10}$ and column $\B\in \mathbb{R}^{k_2\times 10}$ subspaces, fixed sample size defined by $k_1$ and $k_2$ but different set of samples $\widehat{\O}$ drawn without replacement and signal dimension $\Y \in \mathbb{R}^{100\times 100}$. }
\label{fig:projection}
\end{figure}
For completeness, we carry out analysis on more general results, that is on Theorem \ref{th:union}, as all the properties which holds for the general case of missing signals are all applicable for the restricted problem as well. All the parameters $\alpha, \beta, \gamma_1$ and $\gamma_2$  depend on $\delta$ and the lower bound of the Theorem \ref{th:union} contains all these parameters. In order to get more information about the lower bound we set all these parameters very near to $0$, therefore 
% the lower bound approximately equals to 
% \begin{multline*}
% \frac{k_1m_2+k_2m_1-k_1k_2}{m_1m_2}\left(1- \frac{n_1m_2 \mu(\A)+n_2m_1\mu(\B)}{2k_1k_2}  \right) \\\left|\left|\Y- \U^A\Y\U^B\right|\right|^2_F
% \end{multline*}
for a incoherent row and column subspace, that is for $\mu(\A) = 1$ and $\mu(\B) = 1$ we write the lower bound as:
\begin{equation*}
\frac{k_1m_2+k_2m_1-k_1k_2}{m_1m_2}\left(1- \frac{n_1m_2 +n_2m_1}{2k_1k_2}  \right) \\ \left|\left|\Y- \U^A\Y\U^B\right|\right|^2_F.
\end{equation*}
For $k_1<n_1$ and $k_2<n_2$, the first term in the above expression $\frac{k_1m_2+k_2m_1-k_1k_2}{m_1m_2} > 0$. It is because that the actual signal dimension $m_1$ and $m_2$ is greater than the corresponding row and column subspace dimensions $n_1$ and $n_2$. Therefore for $m_1=m_2 = m$, $n_1=n_2=n$ and $k_1=k_2=k$ we write the lower bound as:
\begin{equation*}
\frac{2k(m-k)}{m^2}\left(1- \frac{nm}{k^2}  \right) \left|\left|\Y- \U^A\Y\U^B\right|\right|^2_F.
\end{equation*}
Here the second term is always $\leq 0$ as $nm>k^2$. Therefore, for the incoherent row and column subspaces the lower bound is $\leq 0$, which is consistent with the fact that $\left|\left|\Y_{\Omega} - \U^A_{\Omega}\Y_{\Omega}\U^B_{\Omega} \right|\right|_F^2= 0$ for $k_1<n_1$ and $k_2<n_2$.

For the experiments, we choose highly incoherent subspaces. Both row and column subspaces have Gaussian random bases, that is $\mu(\A) \approx 1.3$ and $\mu(\B) \approx 1.3$. In all these simulations in Fig. \ref{fig:projection} we plot the residual energy $\left|\left|\Y_{\widehat{\O}} - \U^A_{\widehat{\O}}\Y_{\widehat{\O}}\U^B_{\widehat{\O}} \right|\right|_F^2$ as a function of $k_1\cdot k_2$. The plots show the maximum, minimum and mean values of the calculated residual energy over 1000 simulations of $\widehat{\O}$ without replacement for fixed row and column subspace, fixed unit norm signal and $\Y \in (\A \otimes \B)^{\bot}$ in Fig. \ref{fig::ProjectionAB}. We find that the residual energy is always positive when $k_1 \geq n_1\mu(\A)\log\left(n_1\right) $ and $k_2\geq n_2\mu(\B)\log\left(n_2\right)$. The threshold in the Fig. \ref{fig:projection} is calculated as the product of lower bounds of $k_1$ and $k_2$. In Fig. \ref{fig::projectionA}, \ref{fig::projectionB} we also show that the residual energy is still positive when  the signal is sampled from a Kronecker subspace with any one of the subspaces is orthogonal. As expected, the residual energy is $0$ for the signal sampled from the Kronecker subspace itself, that is for $\Y \in (\A \otimes \B)$ the residual energy is $0$, as shown in Fig. \ref{fig::ptojectionNO}.

%!Tex root = Tensor_journal.tex

\section{Matched Subspace Detection}
\label{sec:msd}

In the noise less case $\Y = \A\X\B^T$ we can assume $\eta=0$; we form the detection setup for the hypotheses $\mathcal{H}_0: \Y \in \D$ and $\mathcal{H}_1: \Y \notin \D$. We use the following test statistics
\begin{equation}
\left|\left|\Y_\O - \U_\O^A\Y_\O\U_\O^B\right|\right|_F^2 \underset{\mathcal{H}_1}{\overset{\mathcal{H}_0}{\lessgtr}} \eta.
\end{equation}

For the large enough values of $k_1$ and $k_2$ and $\delta > 0$ the probability of detection is greater than $1-8\delta$, that is \linebreak $\mathbb{P}\left[\left|\left|\Y_\O - \U_\O^A\Y_\O\U_\O^B\right|\right|_F^2 > 0 \Big| \mathcal{H}_1\right]  \geq 1-8\delta$. Also, as shown in Fig. \ref{fig::ptojectionNO}, the projection error is zero when the signal belongs to the Kronecker-structured subspace. Therefore, the probability of false alarm is also 0, $\mathbb{P}\left[\left|\left|\Y_\O - \U_\O^A\Y_\O\U_\O^B\right|\right|_F^2 > 0 \Big| \mathcal{H}_0\right]  =0 $.

With the noisy signal $\widetilde{\Y}= \A\X\B^T + \mathbf{Z}$, where $\mathbf{Z} \sim \mathcal{N}(0, \mathbb{I})$, we use the same hypothesis but the test statistics changes to 
\begin{equation}
\left|\left|\widetilde{\Y}_\O - \U_\O^A\widetilde{\Y}_\O\U_\O^B\right|\right|_F^2 \underset{\mathcal{H}_1}{\overset{\mathcal{H}_0}{\lessgtr}} \eta.
\end{equation}
Here, we note that according to \cite{scharf1994matched}, the test statistics is distributed as a non-central $\mathcal{X}^2$ with $n_1n_2$ degrees of freedom and the non-centrality parameter $\left|\left|\Y_\O - \U_\O^A\Y_\O\U_\O^B\right|\right|_F^2$, Where the detection probability \linebreak ~$P_D = 1 - \mathbb{P}~\left[\mathcal{X}^2_{n_1n_2}\left(\left|\left|\Y_\O - \U_\O^A\Y_\O\U_\O^B\right|\right|_F^2\right) \leq \eta\right]$~ increases  monotonically with the non-centrality parameter. Therefore, this means that the detection probability grows with either $k_1$ or $k_2$ or both.

\section{Conclusion}
\label{sec:conclusion}
In this paper, we extend the matched subspace detection for vectorized signals to tensor signals by projecting the signal onto Kronecker-structured subspace. We have further shown that the detection from a highly incomplete tensor signal is possible by computing the energy outside the KS subspace.

\bibliographystyle{IEEEbib}
\bibliography{subspace_ref_J}

% if have a single appendix:
%\appendix[Proof of the Zonklar Equations]
% or
%\appendix  % for no appendix heading
% do not use \section anymore after \appendix, only \section*
% is possibly needed

% use appendices with more than one appendix
% then use \section to start each appendix
% you must declare a \section before using any
% \subsection or using \label (\appendices by itself
% starts a section numbered zero.)
%

\appendices
%!Tex root = Tensor_journal.tex

\section{Concentration Inequalities}
\label{Appendix:CI}

\begin{theorem}[McDiarmid's Inequality \cite{mcdiarmid141method}] Let $X_1, X_2, \cdots, X_{m_1m_2}$ be the independent random variables and assume a function $f$, such that $f(X_1, X_2, \cdots, X_{m_1m_2}) := Y$, then for $\epsilon > 0$ there exists $t_i$ such that 

\begin{equation}
\mathbb{P}\left[Y\geq E[Y]+\epsilon\right] \leq \exp\left(\frac{-2\epsilon^2}{\sum_{i=1}^{m_1m_2}t_i^2}\right),
\label{eq::mcd_upper}
\end{equation}
\begin{equation}
\mathbb{P}\left[Y\leq E[Y]-\epsilon\right] \leq \exp\left(\frac{-2\epsilon^2}{\sum_{i=1}^{m_1m_2}t_i^2}\right).
\label{eq::mcd_lower}
\end{equation}

\label{th:mcd}
\end{theorem}

\begin{theorem}[Bernstein's Inequality \cite{balzano2010high}] Let $\X_1, \X_2, \cdots, \X_m$ be the independent non-zero $r\times r$ matrices and if for almost all $k$, $\rho_k^2 = \max\{||E[\X_k\X_k^T]||_2, ||E[\X_k^T\X_k]||_2\}$ and $||\X_k||_2 \leq M$ hold true, then for any $\tau > 0$,
\begin{equation}
\mathbb{P}\left[\left|\left|\sum_{i=1}^m \X_i\right|\right|_2 > \tau\right] \leq 2r\exp\left(\frac{-\tau^2/2}{\sum_{i=1}^m\rho_k^2 + M\tau/3}\right)
\end{equation}
\label{th:Bernstein}
\end{theorem}

\section{KS Coherence}
\label{Appendix:coherence}
\begin{lemma}
The coherence of the Kronecker-structured subspace ${\A\otimes \B \in \mathbb{R}^{m_1m_2\times n_1n_2}}$ is equal to the product of coherence of individual subspaces $\A \in \mathbf{R}^{m_1\times n_1}$ and $\B \in \mathbf{R}^{m_2\times n_2}$, that is 
\begin{equation}
\mu(\A \otimes \B) = \mu(\A) \mu(\B).
\label{eq:appcoherenceKS}
\end{equation} 
\label{lm:appcoherence}
\end{lemma}

\begin{IEEEproof}
We write the coherence of Kronecker product of two subspaces as 
\begin{align} 
\mu(\A \otimes \B) = \frac{m_1m_2}{n_1n_2} \underset{j}{\text{max}} \left|\left|\U^{(\A\otimes \B)} e^{AB}_j\right|\right|^2_2,
\label{eq:lemma1}
\end{align}
where $\U^{(\A\otimes \B)} = (\A \otimes \B) \left((\A \otimes \B)^T (\A \otimes \B)\right)^{-1}(\A \otimes \B)^T$ is the orthogonal projector operator onto the Kronecker structured subspace $(\A \otimes \B)$. We can use Kronecker product properties to further simplify this, as
\begin{align*}
\U^{(\A\otimes \B)} &= (\A \otimes \B) \left((\A \otimes \B)^T (\A \otimes \B)\right)^{-1}(\A \otimes \B)^T\\
& = (\A \otimes \B) \left(\A^T\A \otimes \B^T\B \right)^{-1}(\A \otimes \B)^T\\
& = (\A(\A^T\A)^{-1}\A^T) \otimes (\B(\B^T\B)^{-1}\B^T)\\
& = \U^A \otimes \U^B.
\end{align*}
Therefore, we can write \eqref{eq:lemma1} as:
\begin{align}
\mu(\A \otimes \B) &= \frac{m_1m_2}{n_1n_2} \underset{j}{\text{max}} \left|\left|(\U^A \otimes \U^B) e^{AB}_j\right|\right|^2_2.
\end{align}
From \cite{lancaster1972norms}, for the standard basis $e^A_j$ and $e^B_j$ with fixed degrees of freedom, we can always write $e^{AB}_j = e^A_j \otimes e^B_j$ . Thus,
\begin{align}
\mu(\A \otimes \B) &= \frac{m_1m_2}{n_1n_2} \underset{j}{\text{max}} \left|\left|\U^Ae^A_j \otimes \U^Be^B_j\right|\right|^2_2.
\end{align}
We can further simplify the above equation using norm of Kronecker product property in \cite{bernstein2005matrix}. that is, $||\A \otimes \B||_2 = ||\A||_2||\B||_2$. Finally, we can write 
\begin{align}
\mu(\A \otimes \B) &= \frac{m_1m_2}{n_1n_2} \underset{j}{\text{max}} \left(\left|\left|\U^Ae^A_j\right|\right|_2^2 \left|\left| \U^Be^B_j\right|\right|^2_2\right) \nonumber\\
&= \frac{m_1m_2}{n_1n_2} \left(\underset{j}{\text{max}} \left|\left|\U^Ae^A_j\right|\right|_2^2\right) \left(\underset{j}{\text{max}}\left|\left| \U^Be^B_j\right|\right|^2_2\right)\nonumber \\
&= \left(\frac{m_1}{n_1} \underset{j}{\text{max}} \left|\left|\U^Ae^A_j\right|\right|_2^2\right) \left(\frac{m_2}{n_2}\underset{j}{\text{max}}\left|\left| \U^Be^B_j\right|\right|^2_2\right) \nonumber \\
&=\mu(\A)\mu(\B).
\end{align}
\end{IEEEproof}

\section{Signal residual energy}
\begin{IEEEproof}[Proof of lemma \ref{lm::KSEnBound_inter}] To prove this lemma we use Theorem \ref{th:mcd} in Appendix \ref{Appendix:CI}. Let $|\Omega_A| = k_1$ and $|\Omega_B| = k_2$ denotes the number of non-zero rows and columns. Therefore $A_{\Omega} \in \mathbb{R}^{k_1 \times n_1}$ and $B_{\Omega} \in \mathbb{R}^{k_2 \times n_2}$ and one can write $\left|\left|\Y\right|\right|_F^2 = \frac{1}{2}\left(\sum_{i=1}^{m_1}\left|\left|\Y(i,:)\right|\right|_2^2 + \sum_{j=1}^{m_2}\left|\left|\Y(:,j)\right|\right|_2^2\right)$. Set $x_{i,j} = \Y^2_{\O(i,j)}$ and we know that $\Y^2_{\O(i,j)} \leq ||\Y||_{\infty}^2$ is true $\forall i \in \{1, 2, \cdots, k_1\}$ and $\forall j\in \{1,2,\cdots,k_2\}$. Therefore,
%  & = E\left[\left|\left|\Y_{\Omega}\right|\right|_F^2\right]
\begin{align}
E\left[\sum_{i=1}^{k_1}\sum_{j=1}^{k_2}x_{i,j}\right] & = E\left[||\Y||_F^2\right] - E\left[\sum_{i=1}^{m_1-k_1}||\Y_{\Omega}(i,:)||_2^2 + \sum_{j=1}^{m_2-k_2}||\Y_{\Omega}(:,j)||_2^2 - \sum_{i=1}^{m_1-k_1}\sum_{j=1}^{m_2-k_2}(\Y_{\Omega}(i,j))^2\right] \nonumber\\
&= ||\Y||_F^2 - \Bigg( \sum_{i=1}^{m_1-k_1} E\left[\sum_{l=1}^{m_1}||\Y(i,:)||_2^2 \mathbb{1}_{\Omega(i,:) = l} \right] + \sum_{j=1}^{m_2-k_2}E\left[\sum_{l=1}^{m_2}||\Y(:,j)||_2^2 \mathbb{1}_{\Omega(:,j) = l} \right] \nonumber\\& \quad- \sum_{i=1}^{k_1}\sum_{j=1}^{k_2}E\left[\sum_{r=1}^{m_1}\sum_{s=1}^{m_2}(\Y_{\Omega}(i,j))^2\mathbb{1}_{\Omega(i,j) = (r,s)}\right] \Bigg)\nonumber\\
&= ||\Y||_F^2 -\left[ \frac{m_1-k_1}{m_1} ||\Y||_F^2 + \frac{m_2-k_2}{m_2} ||\Y||_F^2 - \frac{(m_1-k_1)(m_2-k_2)}{m_1m_2}||\Y||_F^2\right]\nonumber\\
&=\left[1-\left(\frac{m_1-k_1}{m_1} + \frac{m_2-k_2}{m_2} - \frac{(m_1-k_1)(m_2-k_2)}{m_1m_2}\right)\right]||\Y||_F^2 \nonumber\\
&=\left(\frac{k_1k_2}{m_1m_2}\right)||\Y||_F^2,
\end{align}
where $\mathbb{1}$ represents the indicator function and $\O$ assumes that the samples are taken uniformly without replacement. Thus, from we write the left hand side of \eqref{eq::mcd_lower} as 
\begin{equation*}
\mathbb{P}\left[\left|\sum_{i=1}^{k_1}\sum_{j=1}^{k_2}x_{i,j}\right| \leq E\left[\sum_{i=1}^{k_1}\sum_{j=1}^{k_2}x_{i,j}\right] - \epsilon\right] = \mathbb{P}\left[\left|\sum_{i=1}^{k_1}\sum_{j=1}^{k_2}x_{i,j}\right| \leq \left( \frac{k_1k_2}{m_1m_2} \right)\left|\left|\Y\right|\right|_F^2 - \epsilon\right].
\end{equation*}
For $\epsilon =  \alpha \left( \frac{k_1k_2}{m_1m_2} \right)\left|\left|\Y\right|\right|_F^2$, we can bound this probability by
\begin{align*}
 2\exp\left(\frac{-2\left(\alpha \left(\frac{k_1k_2}{m_1m_2} \right)\left|\left|\Y\right|\right|_F^2\right)^2}{\sum_{i=1}^{k_1k_2}\left(2||\Y||_{\infty}^2\right)^2}\right) &=  2\exp\left(\frac{-2\alpha^2 \left(\frac{k_1k_2}{m_1m_2} \right)^2\left|\left|\Y\right|\right|_F^4}{4k_1k_2||\Y||_{\infty}^4}\right)\\
 &=2\exp\left(\frac{-\alpha^2 \left( k_1k_2 \right)}{2} \left(\frac{\left|\left|\Y\right|\right|_F^2}{m_1m_2||\Y||_{\infty}^2}\right)^2\right)\\
  &=2\exp\left(\frac{-\alpha^2 \left(k_1k_2 \right)}{2\mu(\Y)^2}\right).
\end{align*}
The final bound obtained is:
\begin{equation*}
\mathbb{P}\left[\Y^2_{\O(i,j)} \geq (1-\alpha)\left(\frac{k_1k_2}{m_1m_2} \right)\left|\left|\Y\right|\right|_F^2\right] \geq 1-2\exp\left(\frac{-\alpha^2 \left(k_1k_2 \right)}{2 \mu(\Y)^2}\right).
\end{equation*}
Then for $\alpha = \sqrt{\frac{2\mu(\Y)^2}{k_1k_2}\log(\frac{1}{\delta})}$, with the probability $1-2\delta$:
\begin{equation}
(1-\alpha)\left(\frac{k_1k_2}{m_1m_2}\right)||\Y||_F^2 \leq ||\Y_{\Omega}||_F^2 \\ \leq (1+\alpha)\left(\frac{k_1k_2}{m_1m_2}\right)||\Y||_F^2.
\label{eq::KSEnBound_inter}
\end{equation}

\end{IEEEproof}

\begin{IEEEproof}[Proof of lemma \ref{lm::CoeffBound_inter}]
To prove this lemma we again use Theorem \ref{th:mcd} in Appendix \ref{Appendix:CI} and Lemma \ref{lm:appcoherence} in Appendix \ref{Appendix:coherence}. Let $\mathbf{V}_{\Omega_{i,j}} = \A_{\Omega_{i,:}}^T \Y_{\Omega_{i,j}} \B_{\Omega_{:,j}}$ represent the $(i,j)^{th}$ samples index, where ${\Omega_{i,:}}$ represents the $i^{th}$ row of the column subspace $\A$ and ${\Omega_{:,j}}$ represents the $j^{th}$ row of the row subspace $\B$. Here $\left|\left|\sum_{i=1}^{k_1}\sum_{j=1}^{k_2}\mathbf{V}_{\Omega_{i,j}}\right|\right|_F = \left|\left|\A_{\Omega}^T \Y_{\Omega} \B_{\Omega}\right|\right|_F$. Now, in order to bound $||V||$ we use the vectorized form of the signal and write $\left|\left|\A_{\Omega}^T \Y_{\Omega} \B_{\Omega}\right|\right|_F = ||\mathbf{v}||_2 = \left|\left|(\A\otimes \B)_{\O} \mathbf{y}\right|\right|_2 = ||\D_{\O} \mathbf{y}||_2$, where $\mathbf{y} = \text{vec}(\Y)$. By definition in \eqref{eq:coherence}, we know $||\D_{\O(i)}||_2 = ||\D^Te_i||_2 = \left|\left|\U^{\D} e^{AB}_j\right|\right|_2 \leq \sqrt{\frac{n_1n_2}{m_1m_2}\mu(\D)}$. Now from Lemma \ref{lm:appcoherence} we prove that the coherence of Kronecker subspace is product of coherence of individual subspaces. Therefore, we write $||\D_{\O(i)}||_2 \leq \sqrt{\frac{n_1n_2}{m_1m_2}\mu(\A)\mu(\B)}$. Thus, in vectorized form we write
\begin{equation*}
||\mathbf{v}_i||_2 \leq |\mathbf{y}_{\O(i)}| ||\D_{\O(i)}||_2 \leq ||\Y||_{\infty} \sqrt{\frac{n_1n_2}{m_1m_2}\mu(\A)\mu(\B)}.
\end{equation*}
Now suppose that the samples are take uniformly without replacement, then we can write the following bound as
\begin{align}
 E\left[ \left|\left| \sum_{i=1}^{k_1} \sum_{j=1}^{k_2} \mathbf{V}_{\Omega_{i,j}}\right|\right|_F^2\right] &=E\left[ \left|\left| \sum_{i=1}^{k_1} \sum_{j=1}^{k_2} \A_{\Omega_{i,:}}^T \Y_{\Omega_{i,j}} \B_{\Omega_{:,j}}\right|\right|_F^2\right] \nonumber\\
&= \sum_{p=1}^{n_1}\sum_{q=1}^{n_2} E\left[ \sum_{i=1}^{k_1} \sum_{j=1}^{k_2} \sum_{r=1}^{m_1}\sum_{s=1}^{m_2} \A^2_{rp} \Y^2_{rs} (\mathbb{1}_{r=i, s=j}) B^2_{sq}\right]\nonumber\\
&= \sum_{p=1}^{n_1}\sum_{q=1}^{n_2} \frac{k_1k_2}{m_1m_2} \left(\sum_{r=1}^{m_1}\sum_{s=1}^{m_2} \A^2_{rp} \Y^2_{rs} \B^2_{sq}\right)\nonumber\\
&= \sum_{p=1}^{n_1}\sum_{q=1}^{n_2} \frac{1}{2}\frac{k_1k_2}{m_1m_2} \left(\sum_{r=1}^{m_1}\sum_{s=1}^{m_2} \A^2_{rp} \Y^2_{rs}  + \sum_{r=1}^{m_1}\sum_{s=1}^{m_2} \Y^2_{rs} \B^2_{sq}\right)\nonumber\\ 
&\leq \frac{k_1k_2}{2m_1m_2} \left( \frac{n_1}{m_1} \mu(\A)+  \frac{n_2}{m_2} \mu(\B)\right) ||\Y||^2_F.
\end{align}

Using the McDiard's concentration inequality from Theorem \ref{th:mcd}, we write the left hand side of \eqref{eq::mcd_upper} as 
\begin{multline*}
\mathbb{P}\left[\left|\left| \sum_{i=1}^{k_1} \sum_{j=1}^{k_2} \mathbf{V}_{\Omega_{i,j}}\right|\right|_F^2 \geq E\left[ \left|\left| \sum_{i=1}^{k_1} \sum_{j=1}^{k_2} \mathbf{V}_{\Omega_{i,j}}\right|\right|_F^2\right]+ \epsilon\right] = \\ \mathbb{P}\left[\left|\left| \sum_{i=1}^{k_1} \sum_{j=1}^{k_2} \mathbf{V}_{\Omega_{i,j}}\right|\right|_F^2 \geq \frac{k_1k_2}{2m_1m_2} \left( \frac{n_1}{m_1} \mu(\A)+  \frac{n_2}{m_2} \mu(\B)\right) ||\Y||^2_F + \epsilon\right].
\end{multline*}
For $\epsilon =  \beta \sqrt{\frac{k_1k_2}{2m_1m_2} \left( \frac{n_1}{m_1} \mu(\A)+  \frac{n_2}{m_2} \mu(\B)\right)} ||\Y||_F$, we can bound this probability by
\begin{align*}
 2\exp\left(\frac{-2\left(\beta \sqrt{\frac{k_1k_2}{2m_1m_2} \left( \frac{n_1}{m_1} \mu(\A)+  \frac{n_2}{m_2} \mu(\B)\right)} ||\Y||_F\right)^2}{\sum_{i=1}^{k_1k_2}\left(2||\Y||_{\infty} \sqrt{\frac{n_1n_2}{m_1m_2}\mu(\A)\mu(\B)}\right)^2}\right) &=  2\exp\left(\frac{-\beta^2 (\frac{k_1k_2}{m_1m_2} \left( \frac{n_1}{m_1} \mu(\A)+  \frac{n_2}{m_2} \mu(\B)\right))\left|\left|\Y\right|\right|_F^2}{4k_1k_2||\Y||_{\infty}^2(\frac{n_1n_2}{m_1m_2}\mu(\A)\mu(\B))}\right)\\
 &=2\exp\left(\frac{-\beta^2}{4} \left(\frac{m_2}{n_2}\frac{1}{\mu(\B)}+\frac{m_1}{n_2}\frac{1}{\mu(\A)}\right)\left(\frac{\left|\left|\Y\right|\right|_F^2}{m_1m_2||\Y||_{\infty}^2}\right)\right)\\
  &=2\exp\left(\frac{-\beta^2}{4\mu(\Y)} \left(\frac{m_2}{n_2}\frac{1}{\mu(\B)}+\frac{m_1}{n_2}\frac{1}{\mu(\A)}\right)\right).
\end{align*}
The final bound obtained is:
\begin{multline*}
\mathbb{P}\left[\left|\left| \sum_{i=1}^{k_1} \sum_{j=1}^{k_2} \mathbf{V}_{\Omega_{i,j}}\right|\right|_F^2 \geq (\beta+1)^2\frac{k_1k_2}{2m_1m_2} \left( \frac{n_1}{m_1} \mu(\A)+  \frac{n_2}{m_2} \mu(\B)\right) ||\Y||^2_F\right]  \leq 2\exp\left(\frac{-\beta^2}{4\mu(\Y)} \left(\frac{m_2}{n_2}\frac{1}{\mu(\B)}+\frac{m_1}{n_2}\frac{1}{\mu(\A)}\right)\right).
\end{multline*}
Then for $\beta = \sqrt{\frac{4\mu(\Y)\log(\frac{1}{\delta})}{\left(\frac{m_2}{n_2}\frac{1}{\mu(\B)}+\frac{m_1}{n_2}\frac{1}{\mu(\A)}\right)} }$, with the probability $1-2\delta$:
\begin{equation}
||A_{\Omega}^T Y_{\Omega} B_{\Omega}||_F^2 \leq (\beta+1)^2 \\ \frac{1}{2}\frac{k_1k_2}{m_1m_2} \left( \frac{n_1}{m_1} \mu(\A) +  \frac{n_2}{m_2} \mu(\B)\right) ||\Y||^2_F.
\label{eq::CoeffBound_inter}
\end{equation}

\end{IEEEproof}

\begin{IEEEproof}[Proof of lemma \ref{lm::ABsubBound_inter}] Following the same arguments as in \cite{balzano2010high} and applying the Theorem \ref{th:Bernstein}, we bound the following probability

\begin{equation*}
 P\left[\left|\left|\sum_{i\in \Omega}\left(\A_{\Omega_{i,:}} \A_{\Omega_{i,:}}^T - \frac{1}{m_1} \mathbf{1}_{n_1} \right)\right|\right|_F \leq \frac{k_1}{m_1}\gamma_1 \right] \geq 1-\delta,
 \end{equation*}
with probability $1-\delta$.  We finally bound the energy of row and column subspaces with probability $1-\delta$ as follows
 \begin{equation}
 \left|\left|\left(\sum_{i\in \Omega}\A_{\Omega_{i,:}} \A_{\Omega_{i,:}}^T\right)^{-1}\right|\right|_F \leq \frac{m_1}{(1-\gamma_1)k_1},
 \label{eq::AsubBound}
\end{equation}
\begin{equation}
 \left|\left|\left(\sum_{i\in \Omega}\B_{\Omega_{:,i}} \B_{\Omega_{:,i}}^T\right)^{-1}\right|\right|_F \leq \frac{m_2}{(1-\gamma_2)k_2}.
 \label{eq::BsubBound}
 \end{equation}
Combining \eqref{eq::KSEnBound_inter},\eqref{eq::CoeffBound_inter},\eqref{eq::AsubBound} and \eqref{eq::BsubBound}, we obtain the desired bounds in Theorem \ref{th:intersection} as 

%  \textbf{U.B}\\
%  \begin{equation*}
%  \left|\left|\Y_{\Omega} - \U^A_{\Omega}\Y_{\Omega}\U^B_{\Omega} \right|\right|_F^2  \leq (1+\alpha)\left(\frac{k_1k_2}{m_1m_2}\right)\left|\left|\Y- \U^A\Y\U^B\right|\right|_F^2
%  \end{equation*}

% \textbf{L.B}\\
% \begin{equation*}
% (1-\alpha)\left(\frac{k_1k_2}{m_1m_2}\right)||\Y||_F^2  - (\beta+1)^2 \\ \frac{1}{2}\frac{k_1k_2}{m_1m_2} \left( \frac{n_1}{m_1} \mu(\A) +  \frac{n_2}{m_2} \mu(\B)\right) ||\Y||^2_F  \frac{m_1}{(1-\gamma_1)k_1} \frac{m_2}{(1-\gamma_2)k_2}\leq ||\Y_{\Omega} - \U^A_{\Omega}\Y_{\Omega}\U^B_{\Omega} ||_F^2 
% \end{equation*}

% \begin{equation*}
% \left((1-\alpha)\left(\frac{k_1k_2}{m_1m_2}\right) - \frac{(\beta+1)^2}{2(1-\gamma_2)(1-\gamma_1)} \left( \frac{n_1}{m_1} \mu(\A)+  \frac{n_2}{m_2} \mu(\B)\right) \right) ||\Y- \U^A\Y\U^B||^2_F \leq ||\Y_{\Omega} - \U^A_{\Omega}\Y_{\Omega}\U^B_{\Omega} ||_F^2  
% \end{equation*}

% The final expression we write is:
\begin{multline}
\Bigg((1-\alpha)\left(\frac{k_1k_2}{m_1m_2}\right) - \frac{(\beta+1)^2}{2(1-\gamma_2)(1-\gamma_1)} \Big( \frac{n_1}{m_1} \mu(\A) +  \frac{n_2}{m_2} \mu(\B)\Big) \Bigg) ||\Y- \U^A\Y\U^B||^2_F \leq ||\Y_{\Omega} - \U^A_{\Omega}\Y_{\Omega}\U^B_{\Omega} ||_F^2 \\ \leq (1+\alpha)\left(\frac{k_1k_2}{m_1m_2}\right)\left|\left|\Y- \U^A\Y\U^B\right|\right|_F^2,
\end{multline}

\end{IEEEproof}

\begin{IEEEproof}[Proof of lemma \ref{lm::KSEnBound_union}]
 To prove this lemma we use Theorem \ref{th:mcd} in Appendix \ref{Appendix:CI}. Let $|\Omega_A| = k_1$ and $|\Omega_B| = k_2$ denotes the number of non-zero rows and columns. Therefore $A_{\Omega} \in \mathbb{R}^{k_1 \times n_1}$ and $B_{\Omega} \in \mathbb{R}^{k_2 \times n_2}$ and one can write $\left|\left|\Y\right|\right|_F^2 = \frac{1}{2}\left(\sum_{i=1}^{m_1}\left|\left|\Y(i,:)\right|\right|_2^2 + \sum_{j=1}^{m_2}\left|\left|\Y(:,j)\right|\right|_2^2\right)$. Set $x_{i,j} = \Y^2_{\O(i,j)}$ and we know that $\Y^2_{\O(i,j)} \leq ||\Y||_{\infty}^2$ is true $\forall i \in \{1, 2, \cdots, k_1\}$ and $\forall j\in \{1,2,\cdots,k_2\}$. Therefore,
%  & = E\left[\left|\left|\Y_{\Omega}\right|\right|_F^2\right]
\begin{align}
E\left[\sum_{i=1}^{k_1}\sum_{j=1}^{k_2}x_{i,j}\right] &= E\left[\sum_{i=1}^{k_1}\left|\left|\Y_{\Omega}(i,:)\right|\right|_2^2 + \sum_{j=1}^{k_2}\left|\left|\Y_{\Omega}(:,j)\right|\right|_2^2 - \sum_{i=1}^{k_1}\sum_{j=1}^{k_2}(\Y_{\Omega}(i,j))^2\right]\nonumber\\
&= \sum_{i=1}^{k_1} E\left[\sum_{l=1}^{m_1}\left|\left|\Y(i,:)\right|\right|_2^2 \mathbb{1}_{\Omega(i,:) = l} \right]  + \sum_{j=1}^{k_2}E\left[\sum_{l=1}^{m_2}\left|\left|\Y(:,j)\right|\right|_2^2 \mathbb{1}_{\Omega(:,j) = l} \right] \nonumber \\ &\quad \quad \quad\quad\quad\quad\quad\quad\quad\quad\quad\quad\quad\quad\quad\quad\quad\quad\quad\quad\quad\quad\quad\quad\quad- \sum_{i=1}^{k_1}\sum_{j=1}^{k_2}E\left[\sum_{r=1}^{m_1}\sum_{s=1}^{m_2}(Y_{\Omega}(i,j))^2\mathbb{1}_{\Omega(i,j) = (r,s)}\right] \nonumber\\
&= \frac{k_1}{m_1} \left|\left|\Y\right|\right|_F^2 + \frac{k_2}{m_2} \left|\left|\Y\right|\right|_F^2 - \frac{k_1k_2}{m_1m_2}\left|\left|\Y\right|\right|_F^2\nonumber\\
&=\left(\frac{k_1}{m_1} + \frac{k_2}{m_2} - \frac{k_1k_2}{m_1m_2} \right)\left|\left|\Y\right|\right|_F^2,
\end{align}
where $\mathbb{1}$ represents the indicator function and $\O$ assumes that the samples are taken uniformly without replacement. Thus, from we write the left hand side of \eqref{eq::mcd_lower} as 
\begin{equation*}
\mathbb{P}\left[\left|\sum_{i=1}^{k_1}\sum_{j=1}^{k_2}x_{i,j}\right| \leq E\left[\sum_{i=1}^{k_1}\sum_{j=1}^{k_2}x_{i,j}\right] - \epsilon\right] = \mathbb{P}\left[\left|\sum_{i=1}^{k_1}\sum_{j=1}^{k_2}x_{i,j}\right| \leq \left(\frac{k_1}{m_1} + \frac{k_2}{m_2} - \frac{k_1k_2}{m_1m_2} \right)\left|\left|\Y\right|\right|_F^2 - \epsilon\right].
\end{equation*}
For $\epsilon =  \alpha \left(\frac{k_1}{m_1} + \frac{k_2}{m_2} - \frac{k_1k_2}{m_1m_2} \right)\left|\left|\Y\right|\right|_F^2$, we can bound this probability by
\begin{align*}
 2\exp\left(\frac{-2\left(\alpha \left(\frac{k_1}{m_1} + \frac{k_2}{m_2} - \frac{k_1k_2}{m_1m_2} \right)\left|\left|\Y\right|\right|_F^2\right)^2}{\sum_{i=1}^{k_1k_2}\left(2||\Y||_{\infty}^2\right)^2}\right) &=  2\exp\left(\frac{-2\alpha^2 \left(\frac{k_1}{m_1} + \frac{k_2}{m_2} - \frac{k_1k_2}{m_1m_2} \right)^2\left|\left|\Y\right|\right|_F^4}{4k_1k_2||\Y||_{\infty}^4}\right)\\
 &=2\exp\left(\frac{-\alpha^2 \left(k_1m_2 + k_2m_1 - k_1k_2 \right)^2}{2k_1k_2} \left(\frac{\left|\left|\Y\right|\right|_F^2}{m_1m_2||\Y||_{\infty}^2}\right)^2\right)\\
  &=2\exp\left(\frac{-\alpha^2 \left(k_1m_2 + k_2m_1 - k_1k_2 \right)^2}{2k_1k_2 \mu(\Y)^2}\right).
\end{align*}
The final bound obtained is:
\begin{equation*}
\mathbb{P}\left[\Y^2_{\O(i,j)} \geq (1-\alpha)\left(\frac{k_1}{m_1} + \frac{k_2}{m_2} - \frac{k_1k_2}{m_1m_2} \right)\left|\left|\Y\right|\right|_F^2\right] \geq 1-2\exp\left(\frac{-\alpha^2 \left(k_1m_2 + k_2m_1 - k_1k_2 \right)^2}{2k_1k_2 \mu(\Y)^2}\right).
\end{equation*}
Then for $\alpha = \sqrt{\frac{2\mu(\Y)^2k_1k_2}{(k_1m_2+k_2m_1-k_1k_2)^2}\log(\frac{1}{\delta})}$, with the probability $1-2\delta$:
\begin{equation}
(1-\alpha)\frac{k_1m_2+k_2m_1-k_1k_2}{m_1m_2}||\Y||_F^2 \leq ||\Y_{\Omega}||_F^2 \\ \leq (1+\alpha)\frac{k_1m_2+k_2m_1-k_1k_2}{m_1m_2}||\Y||_F^2.
\label{eq::KSEnBound}
\end{equation}
% \end{lemma}
\end{IEEEproof}

\begin{IEEEproof}[Proof of lemma \ref{lm::CoeffBound_union}]
To prove this lemma we again use Theorem \ref{th:mcd} in Appendix \ref{Appendix:CI} and Lemma \ref{lm:appcoherence} in Appendix \ref{Appendix:coherence}. Let $\mathbf{V}_{\Omega_{i,j}} = \A_{\Omega_{i,:}}^T \Y_{\Omega_{i,j}} \B_{\Omega_{:,j}}$ represent the $(i,j)^{th}$ samples index, where ${\Omega_{i,:}}$ represents the $i^{th}$ row of the column subspace $\A$ and ${\Omega_{:,j}}$ represents the $j^{th}$ row of the row subspace $\B$. Here $\left|\left|\sum_{i=1}^{k_1}\sum_{j=1}^{k_2}\mathbf{V}_{\Omega_{i,j}}\right|\right|_F = \left|\left|\A_{\Omega}^T \Y_{\Omega} \B_{\Omega}\right|\right|_F$. Now, in order to bound $||V||$ we use the vectorized form of the signal and write $\left|\left|\A_{\Omega}^T \Y_{\Omega} \B_{\Omega}\right|\right|_F = ||\mathbf{v}||_2 = \left|\left|(\A\otimes \B)_{\O} \mathbf{y}\right|\right|_2 = ||\D_{\O} \mathbf{y}||_2$, where $\mathbf{y} = \text{vec}(\Y)$. By definition in \eqref{eq:coherence}, we know $||\D_{\O(i)}||_2 = ||\D^Te_i||_2 = \left|\left|\U^{\D} e^{AB}_j\right|\right|_2 \leq \sqrt{\frac{n_1n_2}{m_1m_2}\mu(\D)}$. Now from Lemma \ref{lm:appcoherence} we prove that the coherence of Kronecker subspace is product of coherence of individual subspaces. Therefore, we write $||\D_{\O(i)}||_2 \leq \sqrt{\frac{n_1n_2}{m_1m_2}\mu(\A)\mu(\B)}$. Thus, in vectorized form we write
\begin{equation*}
||\mathbf{v}_i||_2 \leq |\mathbf{y}_{\O(i)}| ||\D_{\O(i)}||_2 \leq ||\Y||_{\infty} \sqrt{\frac{n_1n_2}{m_1m_2}\mu(\A)\mu(\B)}.
\end{equation*}
Now suppose that the samples are take uniformly without replacement, then we can write the following bound as
\begin{align}
E\left[ \left|\left| \sum_{i=1}^{k_1} \sum_{j=1}^{k_2} \mathbf{V}_{\Omega_{i,j}}\right|\right|_F^2\right] &=E\left[ \left|\left| \sum_{i=1}^{k_1} \sum_{j=1}^{k_2} \A_{\Omega_{i,:}}^T \Y_{\Omega_{i,j}} \B_{\Omega_{:,j}}\right|\right|_F^2\right] \nonumber\\
&= \sum_{p=1}^{n_1}\sum_{q=1}^{n_2} E\left[ \sum_{i=1}^{k_1} \sum_{j=1}^{k_2} \sum_{r=1}^{m_1}\sum_{s=1}^{m_2} \A^2_{rp} \Y^2_{rs} (\mathbb{1}_{r=i, s=j}) \B^2_{sq}\right]\nonumber\\
&= \sum_{p=1}^{n_1}\sum_{q=1}^{n_2} \frac{k_1m_2+k_2m_1-k_1k_2}{m_1m_2} \left(\sum_{r=1}^{m_1}\sum_{s=1}^{m_2} \A^2_{rp} \Y^2_{rs} \B^2_{sq}\right)\nonumber\\
&= \sum_{p=1}^{n_1}\sum_{q=1}^{n_2} \Bigg(\frac{k_1m_2+k_2m_1-k_1k_2}{2m_1m_2} \left(\sum_{r=1}^{m_1}\sum_{s=1}^{m_2} \A^2_{rp} \Y^2_{rs}  + \sum_{r=1}^{m_1}\sum_{s=1}^{m_2} \Y^2_{rs} \B^2_{sq}\right)\Bigg) \nonumber\\ 
&\leq \frac{k_1m_2+k_2m_1-k_1k_2}{2m_1m_2} \left( \frac{n_1}{m_1} \mu(\A)+  \frac{n_2}{m_2} \mu(\B)\right) ||\Y||^2_F.
\end{align}
Using the McDiard's concentration inequality from Theorem \ref{th:mcd}, we write the left hand side of \eqref{eq::mcd_upper} as 
\begin{multline*}
\mathbb{P}\left[\left|\left| \sum_{i=1}^{k_1} \sum_{j=1}^{k_2} \mathbf{V}_{\Omega_{i,j}}\right|\right|_F^2 \geq E\left[ \left|\left| \sum_{i=1}^{k_1} \sum_{j=1}^{k_2} \mathbf{V}_{\Omega_{i,j}}\right|\right|_F^2\right]+ \epsilon\right] = \\ \mathbb{P}\left[\left|\left| \sum_{i=1}^{k_1} \sum_{j=1}^{k_2} \mathbf{V}_{\Omega_{i,j}}\right|\right|_F^2 \geq \frac{k_1m_2+k_2m_1-k_1k_2}{2m_1m_2} \left( \frac{n_1}{m_1} \mu(\A)+  \frac{n_2}{m_2} \mu(\B)\right) ||\Y||^2_F + \epsilon\right].
\end{multline*}
For $\epsilon =  \beta \sqrt{\frac{k_1m_2+k_2m_1-k_1k_2}{2m_1m_2} \left( \frac{n_1}{m_1} \mu(\A)+  \frac{n_2}{m_2} \mu(\B)\right)} ||\Y||_F$, we can bound this probability by
\begin{align*}
 & 2\exp\left(\frac{-2\left(\beta \sqrt{\frac{k_1m_2+k_2m_1-k_1k_2}{2m_1m_2} \left( \frac{n_1}{m_1} \mu(\A)+  \frac{n_2}{m_2} \mu(\B)\right)} ||\Y||_F\right)^2}{\sum_{i=1}^{k_1k_2}\left(2||\Y||_{\infty} \sqrt{\frac{n_1n_2}{m_1m_2}\mu(\A)\mu(\B)}\right)^2}\right) \\
 &=  2\exp\left(\frac{-\beta^2 (\frac{k_1m_2+k_2m_1-k_1k_2}{m_1m_2} \left( \frac{n_1}{m_1} \mu(\A)+  \frac{n_2}{m_2} \mu(\B)\right))\left|\left|\Y\right|\right|_F^2}{4k_1k_2||\Y||_{\infty}^2(\frac{n_1n_2}{m_1m_2}\mu(\A)\mu(\B))}\right)\\
 &=2\exp\left(\frac{-\beta^2}{4} \left(\frac{m_1}{k_1}+\frac{m_2}{k_2}-1\right)\left(\frac{m_2}{n_2}\frac{1}{\mu(\B)}+\frac{m_1}{n_2}\frac{1}{\mu(\A)}\right)\left(\frac{\left|\left|\Y\right|\right|_F^2}{m_1m_2||\Y||_{\infty}^2}\right)\right)\\
  &=2\exp\left(\frac{-\beta^2}{4\mu(\Y)} \left(\frac{m_1}{k_1}+\frac{m_2}{k_2}-1\right)\left(\frac{m_2}{n_2}\frac{1}{\mu(\B)}+\frac{m_1}{n_2}\frac{1}{\mu(\A)}\right)\right).
\end{align*}
The final bound obtained is:
\begin{multline*}
\mathbb{P}\left[\left|\left| \sum_{i=1}^{k_1} \sum_{j=1}^{k_2} \mathbf{V}_{\Omega_{i,j}}\right|\right|_F^2 \geq (\beta+1)^2\frac{k_1m_2+k_2m_1-k_1k_2}{2m_1m_2} \left( \frac{n_1}{m_1} \mu(\A)+  \frac{n_2}{m_2} \mu(\B)\right) ||\Y||^2_F\right] \\ \leq 2\exp\left(\frac{-\beta^2}{4\mu(\Y)} \left(\frac{m_1}{k_1}+\frac{m_2}{k_2}-1\right)\left(\frac{m_2}{n_2}\frac{1}{\mu(\B)}+\frac{m_1}{n_2}\frac{1}{\mu(\A)}\right)\right).
\end{multline*}
Then for $\beta = \sqrt{\frac{4\mu(\Y)\log(\frac{1}{\delta})}{\left(\frac{m_1}{k_1}+\frac{m_2}{k_2}-1\right)\left(\frac{m_2}{n_2}\frac{1}{\mu(\B)}+\frac{m_1}{n_2}\frac{1}{\mu(\A)}\right)} }$, with the probability $1-2\delta$:
\begin{equation}
||\A_{\Omega}^T \Y_{\Omega} \B_{\Omega}||_F^2 \leq (\beta+1)^2\frac{k_1m_2+k_2m_1-k_1k_2}{2m_1m_2} \\ \left( \frac{n_1}{m_1} \mu(\A)+  \frac{n_2}{m_2} \mu(\B)\right) ||\Y||^2_F.
\label{eq::CoeffBound}
\end{equation}
Combining \eqref{eq::KSEnBound},\eqref{eq::CoeffBound},\eqref{eq::AsubBound} and \eqref{eq::BsubBound} we obtain the desired bounds in Theorem \ref{th:union} as 
\begin{multline}
\frac{k_1m_2+k_2m_1-k_1k_2}{m_1m_2}\left((1-\alpha) - \frac{(\beta+1)^2m_1m_2}{2k_1k_2(1-\gamma_1)(1-\gamma_2)}\left( \frac{n_1}{m_1} \mu(\A)+  \frac{n_2}{m_2} \mu(\B)\right) \right)||Y||^2_F   \leq ||\Y_{\Omega} - \U^A_{\Omega}\Y_{\Omega}\U^B_{\Omega} ||_F^2 \\ \leq (1+\alpha)\frac{k_1m_2+k_2m_1-k_1k_2}{m_1m_2}||\Y||_F^2.
\end{multline}

\end{IEEEproof}

\end{document}